\documentclass[useAMS,usenatbib]{mn2e}

\usepackage{aas_macros}
\usepackage{graphicx}
\usepackage{hyperref}

\newcommand{\xrt}{\textit{Swift}/XRT}

\newcommand{\po}{power-law}

\newcommand{\bk}{broken power-law}

\newcommand{\nh}{$N_\mathrm{H}$}
\newcommand{\obj}{S5\,0716+714}

\title[Swift/XRT view on S5 0716+714 during a flare]{Swift/XRT view on S5\,0716+714 during a flare}

\author[A. Wierzcholska and H. Siejkowski]{
    A. Wierzcholska$^{1,2}$\thanks{E-mail: \href{mailto:alicja.wierzcholska@ifj.edu.pl}{alicja.wierzcholska@ifj.edu.pl}} and H. Siejkowski$^{3}$\\
    $^{1}$Insitute of Nuclear Physics, Polish Academy of Sciences, ul. Radzikowskiego 152, 31-342 Krak\'{o}w, Poland\\
    $^{2}$Landessternwarte, Universit\"at Heidelberg, K\"onigstuhl, D 69117 Heidelberg, Germany\\
$^{3}$AGH University of Science and Technology, ACC Cyfronet AGH, ul. Nawojki 11, PO Box 386, 30-950, Krak\'{o}w 23, Poland}

\begin{document}

\date{Accepted 2015 May 13 Received 2015 May 9; in original form 2015 March 29}

\pagerange{\pageref{firstpage}--\pageref{lastpage}} \pubyear{2015}

\maketitle

\label{firstpage}

\begin{abstract}
The blazar \obj\ has been monitored in its flaring state between 2015 January 19 and February 22 with \xrt. During this period an exceptional flux level  was observed in the X-ray range as well as  in the other wavelengths, e.g. optical, near infrared and very high energy $\gamma$~rays.
Here, we report X-ray observations of \obj\ carried out during the outburst. The observed X-ray spectra, well described with \bk\ model, disentangle both synchrotron and inverse Compton components. The break energy shifts towards higher energies with increasing flux, revealing the dominance of synchrotron radiation in the X-ray spectrum observed.
We also report spectrum softening with increasing flux.
During the recent flare, significant temporal intranight variability is observed in the X-ray range.
\end{abstract}


\section{Introduction}
Blazars are the most extreme Active Galactic Nuclei with non-thermal emission observed from the jet visible at small angle to the observer's line of sight \citep[e.g.,][]{begelman84}. This group of sources includes BL Lacertae type objects as well as quasar-type blazars known as flat-spectrum radio quasars (FSRQs), high-polarization quasars (HPQs) or optically violent variables (OVVs), depending on the properties used for classification i.e.: radio spectral index, optical polarization or optical variability, respectively. 
Broad-band emission is observed from radio frequencies up to high energy or very high energy gamma rays 
\citep[e.g.,][]{Wagner2009, Vercellone,  Aharonian13_0301, Abramowski2014}. 
The characteristic feature of blazars is their variability, manifested in all wavelengths, with different variability time scales down to hours or even minutes in the most extreme cases
\citep[e.g.][]{wagner,2155flare,Gopal-Krishna11,saito, Liao15}.

The spectral energy distribution (SED), in $\nu$-$\nu F_{\nu}$ representation has a characteristic double-humped structure. This curvature of SED has been widely discussed in the literature, considering both leptonic and hadronic scenarios as a possible explanation \citep[for review see e.g.][]{Bottcher13}.
In the most popular, so-called Synchrotron-self-Compton scenario, the first bump is attributed to the synchrotron radiation originating from relativistic electrons from the jet, while the second is produced in inverse Compton process, involving the same population of electrons and jet synchrotron photons.
The location of peaks in SED allows to subdivide sources into high-energy, intermediate-energy and low-energy peaked objects: HBL, IBL and LBL, respectively 
\citep[see, e.g.,][]{Abdo2010}.

\obj\ is a very bright
\citep[e.g.][]{Aliu12_0413} 
IBL type blazar  located at $z=0.31$ \citep{Donato01}. 
Several multi-frequency campaigns have targeted this source \citep[e.g.][]{Dai13,Liao14, Wu14}. 
The objects was also frequently monitored in X-ray range with different instruments reporting both spectral and temporal variability \citep[e.g.][]{Wagner92, Cappi94, Liao14}.
It is worth mentioning here that  X-ray observations of \obj\ revealed both synchrotron and inverse Compton components manifested in this domain  \citep{Cappi94,Giommi99,Tagliaferri03,Donato05,Ferrero06}.

\section{Observations and data analysis}

The \textit{Swift} mission \citep{Gehrels04} is a multi-wavelength space observatory equipped with following detectors: the Burst Alert Telescope \citep[BAT,][]{Barthelmy05}, X-ray Telescope \citep[XRT,][]{Burrows05} and Ultraviolet/Optical Telescope \citep[UVOT,][]{Roming05}.
Here,  we study  X-ray observations in the energy range of 0.3-10\,keV obtained with \xrt.
Since April 2005 \obj\ has been monitored in several pointed observations, both in photon counting (PC) and windowed timing (WT) modes. 
In this study, we focus on the  flare observed during the period of 2015 January 19--February 22.
The total exposure of the \xrt\ observations studied (IDs 00035009146-00035009202) is 152\,ks.
Data were analysed using version 6.16 of the \textsc{heasoft} package\footnote{\url{http://heasarc.gsfc.nasa.gov/docs/software/lheasoft}} with \textsc{caldb} v.20140120 following the standard procedure \verb|xrtpipeline|.
Spectral analysis is performed in the energy range of 0.3-10\,keV with the latest version of \textsc{xspec} package (version 12.8.2).
All data are binned to have a minimum of 30 counts per bin.
A single \po\ model and also a \bk\ one are tested, both with  hydrogen Galactic absorption $N_\mathrm{H}=3.22\cdot10^{20}$\,cm$^{-2}$ \citep{Kalberla05} fixed as a frozen parameter.

\begin{figure*}
\centering{\includegraphics[width=0.99\textwidth]{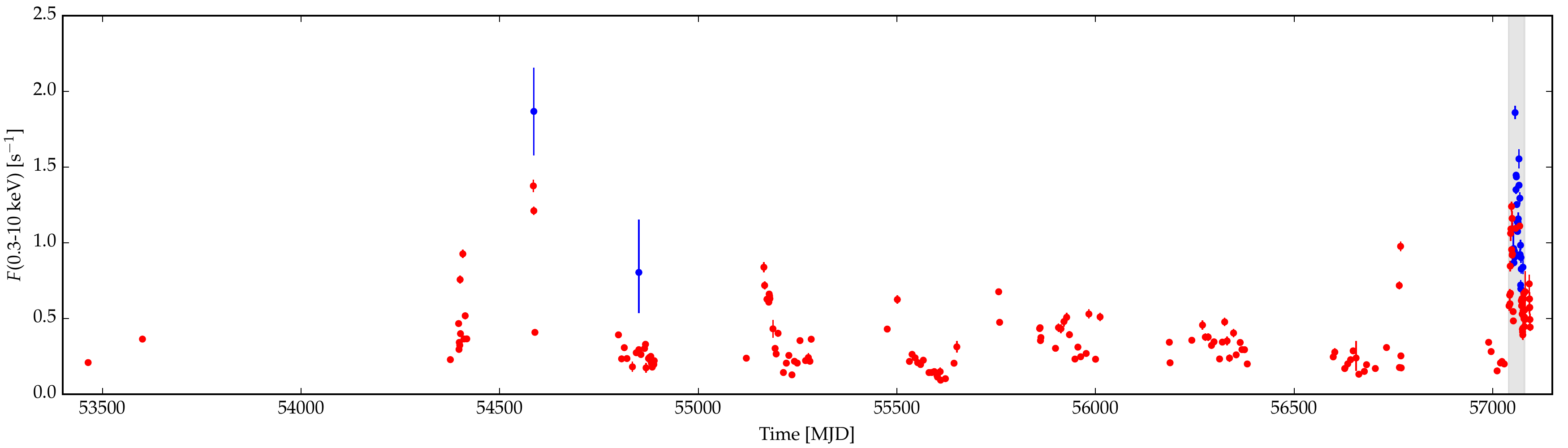}}
\caption{\xrt\ long-term light curve of \obj\ including monitoring of the blazar since April 2005 in the energy range of 0.3-10\,keV. Grey shaded area indicates flare interval discussed in this paper. Observations taken in PC mode and WT mode are marked in red and blue, respectively.}
\label{lc_all}
\end{figure*}

\section{Spectral and temporal variability} \label{variability}
\begin{figure}
\centering{\includegraphics[width=0.48\textwidth]{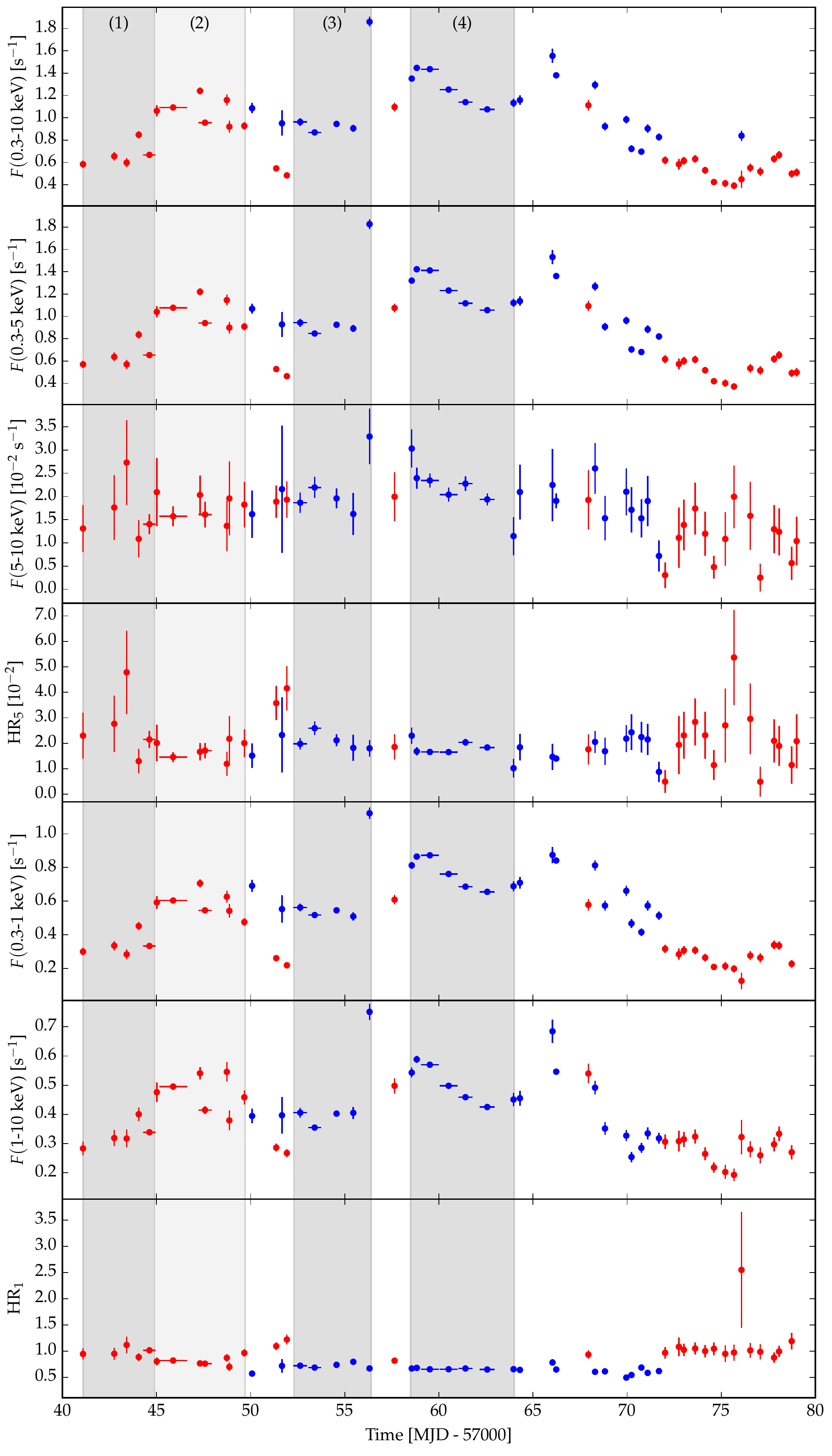}}
\caption{The light curve of \obj\ presenting \xrt\ observations taken during the recent flare. The following panels show light curve in the energy bands of: 0.3-10\,keV, 0.3-5\,keV, 5-10\,keV, 0.3-1\,keV,  1-10\,keV and the corresponding hardness ratios.  The intervals selected for spectral analysis discussed in Sec. \ref{variability} are marked with grey areas. Observations taken in PC mode and WT mode are denoted in red and blue, respectively.
}
\label{lc_short}
\end{figure}

The long-term light curve of \obj\ and a zoom of the flare studied are presented in Fig.~\ref{lc_all} and in the upper panel of Fig.~\ref{lc_short}, respectively. 
During the outburst, significant variability of the source is observed. The maximum flux during this flare is about 1.8\,counts\,s$^{-1}$. It is worth mentioning here that the flare reaches the highest flux level ever observed for \obj\ with \xrt\
and it is best sampled outburst for this source with this instrument.
Data collected during the flare (separately in PC and WT mode) were fitted with single \po\ and \bk\ models, in both cases with Galactic absorption. 
A comparison of the fit parameters for both models, as well as values of the $\chi^2$ statistic, is presented in Table~\ref{table_fits}.
A single \po\ model with a frozen value of \nh\ yields worse fit to data.

\begin{figure}
\centering{\includegraphics[width=0.48\textwidth]{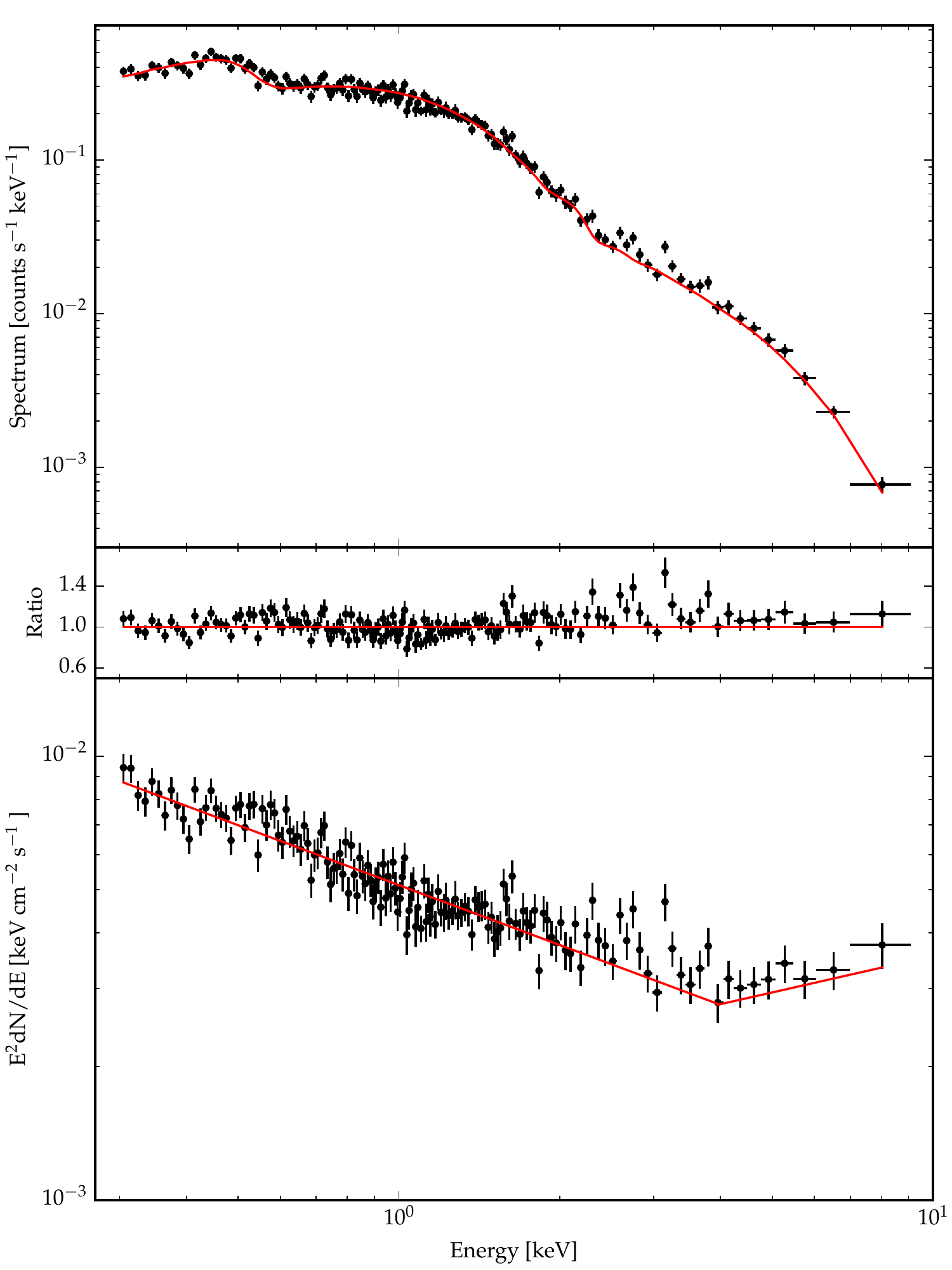}}
\caption{The figure presents spectrum as well as spectral energy distribution of \obj\ including all observations taken during the recent flare in the PC mode. The data are fitted with \bk\ model. The upper panel presents spectral points with fitted model, the middle one shows data and folded-model ratio, while in bottom one the spectral energy distribution is presented. The red line shows the fitted model.}
\label{spec}
\end{figure}

Thus, the \bk\ model is the preferable description of the spectrum for \obj. 
The higher value of $\chi^2_{red}$ in WT mode was expected, since only the PC mode retains full imaging and spectroscopic resolution. In the spectra derived, in PC mode as well as in WT mode, an upturn for \bk\ model is revealed at about 4-5\,keV. Fig.~\ref{spec} shows an example of \bk\ fit to the data in the case of all observations of the recent flare in PC mode. The corresponding spectral energy distribution with an upturn point of 3.98\,keV is presented in the bottom panel of  Fig.~\ref{spec}.

In order to investigate spectral variability in greater detail, the single \po\ and \bk\ models were also tested for four shorter intervals of observations, marked in Fig.~\ref{lc_short} and defined in Table~\ref{table_obsid}. The intervals were optimized to have good enough statistics and observed either in PC or WT mode.
The fit parameters for models as well as values of $\chi^2$ statistics are collected in Table~\ref{table_fits}.
For all the intervals both \po\ as well as \bk\ models were tested using \textit{F}-test \citep[e.g.][]{bevington2003data}, resulting in probability value for predominance of \po\ model of 6, 0.2, 6.6 and $<1$ per cent for intervals (1), (2), (3) and (4), respectively.
Hence, the favourable model of the spectra for these four intervals is a \bk disentangling both synchrotron and inverse Compton components in the X-ray regime. The break energy shifts towards higher values with higher flux levels.

Fig.~\ref{lc_short} shows light curves in five different energy bands, i.e.: 0.3-10, 0.3-5, 5-10, 0.3-1 and 1-10\,keV. As was shown in the case of the integrated spectrum for \obj\, the upturn is located at about 5\,keV. Hence, the light curves in the energy bands of 0.3-5 and 5-10\,keV (the second and the third panels in Fig.~\ref{lc_short}) compare synchrotron and inverse Compton emission, respectively. The light curve in 0.3-5\,keV band follows the one in 0.3-10\,keV energy range. The low statistics of inverse Compton photons cause a worse description of this part of the spectrum (see Table~\ref{table_fits}). This also does not allow us either to claim or to exclude variability of the inverse Compton component. The flux changes in different energy bands at 0.3-10, 0.3-5, 0.3-1 and 1-10\,keV, reveal similar variability patterns and seem to be correlated. 

\begin{figure}
\includegraphics[width=0.48\textwidth]{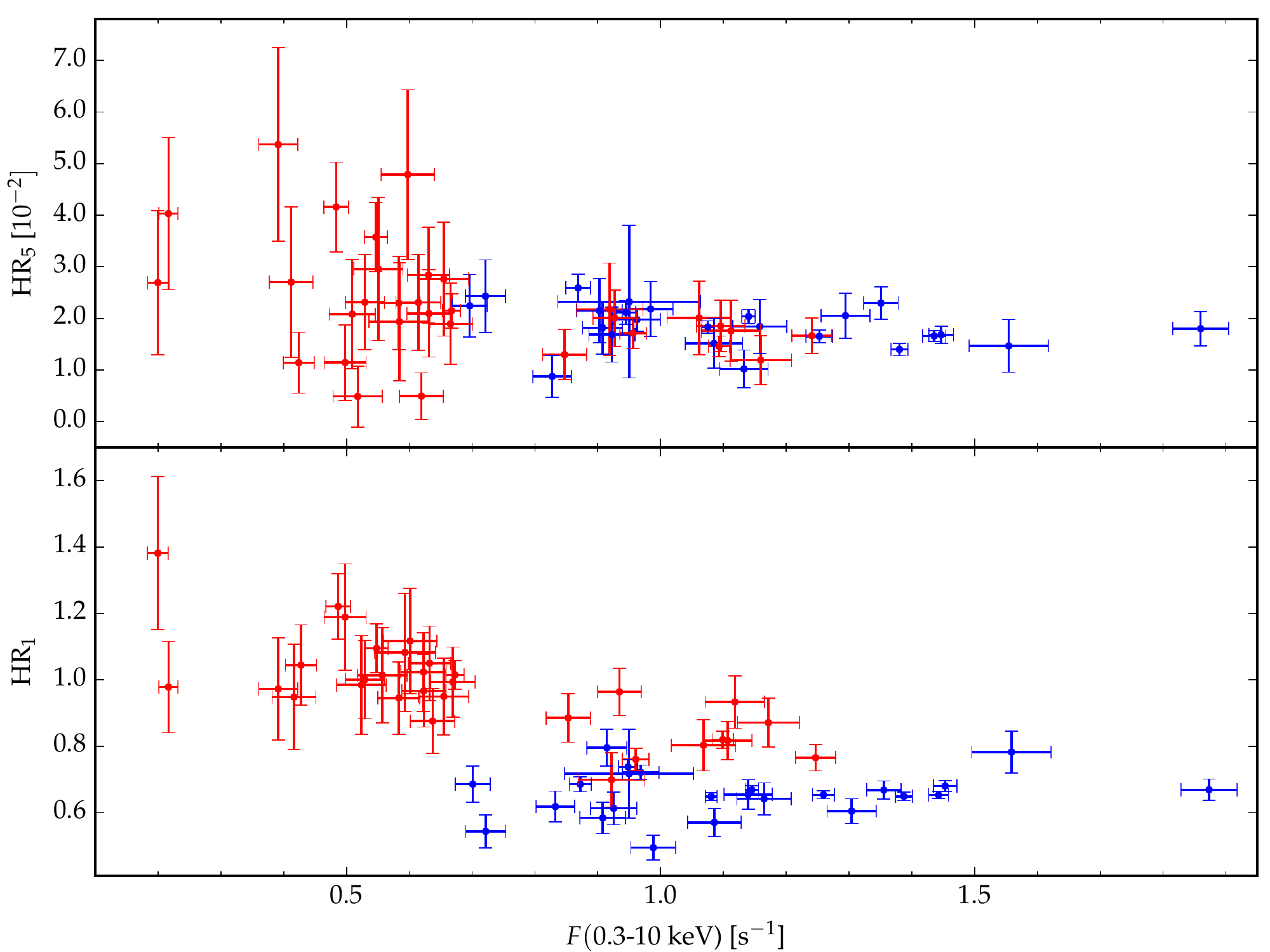}
\caption{The HR$_5 = F($5-10\,keV$)$/$F($0.3-5\,keV$)$ and HR$_1 = F($1-10\,keV$)$/$F($0.3-1\,keV$)$ hardness ratios of \obj\ as a function of total source intensity. Observations taken in PC mode and WT mode are marked in red and blue, respectively.
One point with large error bars is removed from the second panel for better visibility.}
\label{hr_flux}
\end{figure}

The comparison of the hardness ratios for the two cases defined as: HR$_5 = F($5-10\,keV$)$/$F($0.3-5\,keV$)$ and HR$_1 = F($1-10\,keV$)$/$F($0.3-1\,keV$)$, with the source intensity $F($0.3-10\,keV$)$ (see Fig.~\ref{hr_flux}) shows evidence in both cases for spectral steepening with increasing intensity. We denote the flux count rate for a given energy range as $F(E_\textit{min}$, $E_\textit{max})$. The comparison reveals an anticorrelation in the hardness-ratio-intensity plot. This behaviour has been found in several IBL-type blazars \citep[e.g.][]{Ferrero06}, while in the case of HBL objects harder-when-brighter chromatism is observed \citep[e.g.][]{Takahashi96, Takahashi99}. This effect, described by \cite{Padovani96}, is explained in the context of the synchrotron variability, which steepens the overall X-ray spectrum with increasing total flux \citep[e.g.][]{Giommi99, Ferrero06, Foschini06}. 

\begin{table*}  
\caption[]{The parameters for \po\ and \bk\ fits to the data. The following columns give: the number of interval or information about data used; the normalization, the photon indices, the break energy and the reduced $\chi^2$ value for \bk\ fit; the normalization, the photon index and the reduced $\chi^2$ value for \po\ fit. Normalizations N$_{br}$ and N$_{po}$ are given in 10$^{-3}$\,cm$^{-2}$\,s$^{-1}$\,keV$^{-1}$, while break energy is expressed in keV.} 
\centering
\begin{tabular}{c|c|c|c|c|c|c|c|c}
\hline
Interval & N$_{br}$  & $\gamma_1$ & $\gamma_2$ & E$_{br}$ & $\chi^2_{red, br}$(d.o.f.) & N$_{po}$  & $\gamma$ & $\chi^2_{red, po}$(d.o.f.) \\
         \hline
(1)    & 3.913$\pm$0.061  & 2.375$\pm$0.028 & 1.08$\pm$0.10 & 4.68$\pm$0.74 & 1.111(214) & 3.927$\pm$0.061 & 2.353$\pm$0.025 & 1.130(216)  \\
(2)    & 6.466$\pm$0.061  & 2.541$\pm$0.015 & 0.20$\pm$0.80 & 5.75$\pm$0.50 & 0.957(288) & 6.479$\pm$0.061 & 2.532$\pm$0.015 & 0.994(290) \\
(3)    & 5.184$\pm$0.054  & 2.391$\pm$0.018 & 1.27$\pm$0.42 & 5.10$\pm$0.50 & 1.294(263) & 5.187$\pm$0.054 & 2.380$\pm$0.018 & 1.311(265) \\
(4)    & 7.657$\pm$0.029  & 2.663$\pm$0.068 & 0.10$\pm$0.10 & 5.10$\pm$0.50 & 2.473(268) & 7.688$\pm$0.029 & 2.646$\pm$0.066 & 2.862(470) \\[0.5em]
all PC & 5.125$\pm$0.036  & 2.447$\pm$0.012 & 1.73$\pm$0.20 & 3.98$\pm$0.38 & 1.177(390) & 5.150$\pm$0.035 & 2.425$\pm$0.011 & 1.237(392) \\
all WT & 7.050$\pm$0.025  & 2.622$\pm$0.062 & 0.10$\pm$0.10 & 5.20$\pm$0.40 & 2.663(514) & 7.072$\pm$0.025 & 2.606$\pm$0.061 & 3.003(516) \\

\hline
     
\end{tabular}

\label{table_fits}
\end{table*} 

\begin{table}  
\caption[]{Selected intervals for detailed spectral studies. The following columns present: the number of the interval or information about data used; the observation ID\,$-$\,00035009000; the total exposure; the observation mode.} 
\centering
\begin{tabular}{c|c|c|c}
\hline
Interval & Observation IDs & Exposure [ks] & Mode  \\
\hline
(1) & 147--153  & 11.1  & PC    \\
(2) & 154--161 & 23.3  & PC  \\
(3) & 169--173 & 19.3  & WT   \\
(4) & 175--188  & 77.4  & WT     \\[0.5em]
all PC &  147--202 & 102.9 & PC \\
all WT &  147--202 & 49.1 & WT  \\

\hline
     
\end{tabular}

\label{table_obsid}
\end{table}

\begin{figure*}
\centering{\includegraphics[width=0.9\textwidth]{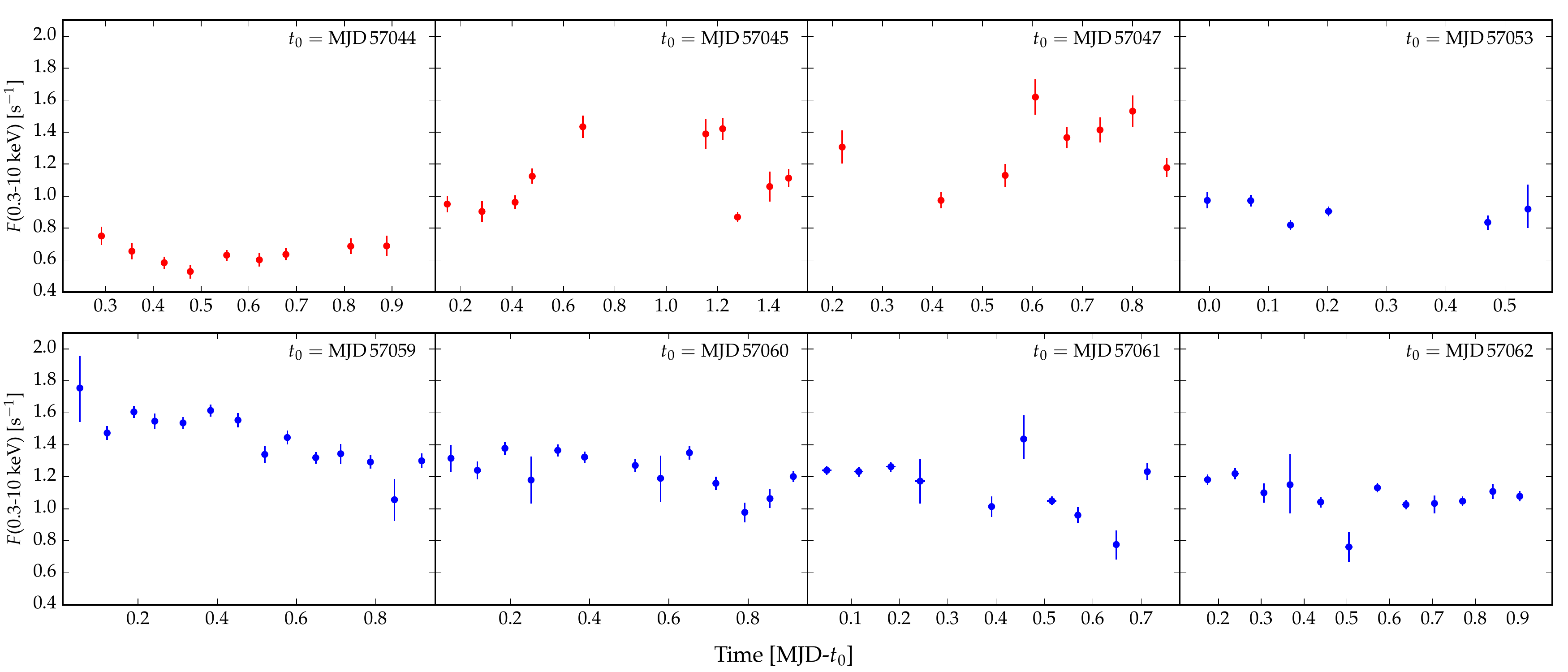}}
\caption{Intra-night variability visible during 8 selected nights of the recent flare. Observations taken in PC mode and WT mode are marked in red and blue, respectively. Modified Julian Date of each observation night is given in right top corner in each plot and the x-axis shows fractional part of the day.}
\label{intranight}
\end{figure*}

The recent intensive monitoring of \obj\ results in eight nights during which the exposure of the observations is larger than 6\,ks. For these nights, detailed light curves (with snapshot-wise bins) are presented in Fig.~\ref{intranight}. The relatively high flux observed in each case gives a perfect possibility of studying intranight variability in the X-ray regime. 
In all the cases significant  variability is detected, fitting with a constant to data points results in a value of $\chi^2_{red}$ of at least 2. The largest intranight variability is observed on MJD\,57054 with $\chi^2_{red}$ equal to 16.

\section{Summary and conclusions}
\xrt\ observations of \obj\ performed during the period of 2015 January 19--February 22 show significant brightness increase of the source. \obj\ is known to be a very variable object, and not only in the X-ray range \citep[e.g.][]{Wagner92, Cappi94,  Tagliaferri03,Donato05, Liao14}. 
The elevated flux of the source during the most recent flare was reported in different wavelengths e.g. optical \citep{ATel6944, ATel6957, ATel7004}, near-infrared \citep{ATel6902},  and very high energy gamma rays \citep{ATel6999}.
During the period observed, the flaring activity was not detected in GeV monitoring with \textit{Fermi} Large Area Telescope\footnote{\url{http://fermi.gsfc.nasa.gov/ssc/data/access/lat/msl_lc/}}.

Since the monitoring programmes mentioned do not provide publicly available light curves, we are not able to judge whether the shape of the flare is similar at other wavelengths or not. However, we note here, that the mentioned Astronomer's Telegrams report an exceptional behaviour of \obj\ starting from January 16. The \xrt\ monitoring of the blazar started on January 19, but the highest flux level observed was detected in the X-ray regime a few days later.

Previous studies of the source in X-ray domain revealed both synchrotron and inverse Compton components in the X-ray range, with break energy in the range of 1.5-3.0\,keV \citep{Tagliaferri03,Donato05,Ferrero06, Foschini06}. We note here that observations mentioned were mostly performed in the quiescence state of the blazar. \cite{Ferrero06} studied changes of the break energy in \bk\ spectral fit for different flux levels. The authors have shown that, during the elevated flux-level period, the crossing point is shifted to a higher energy of about 5\,keV, while for the lower flux states the upturn occurs at about 2\,keV. 

In our studies, we found that the X-ray variable spectrum of \obj\ is well described with \bk\ model. 
This description  is consistent with synchrotron cooling processes associated with a single emitting component. The lack of any significant spectral variability also confirms that there is no need of any additional emitting component. 
The break energy in \bk\ description shifts to higher values with increasing flux level. We found that most of the X-ray emission is observed in the energy range of 0.3-5\,keV.
Furthermore, we found that source follows the softer-when-brighter trend typical for IBL-type objects \citep[e.g.][]{Ferrero06}.

The temporal variability studies of \obj\ have revealed significant intranight variability in the X-ray regime. Previously, in this blazar, such feature has been observed mainly in the optical regime \citep[e.g.][]{Montagni06,Bogdan15}.

\section*{Acknowledgements}
We thank  the Swift PI, Neil Gehrels, for accepting our request for ToO observations of \obj\ with \xrt.
A.W. acknowledges support by Polish Ministry of Science and Higher Education in Mobility Plus Program and  Polish National Science Center for supporting this work  through grant  DEC$-$2011/03/N/ST9/01867.
This research was supported in part by PLGrid Infrastructure.

\bibliographystyle{mn2e_williams}
\bibliography{references}

\begin{thebibliography}{40}
\expandafter\ifx\csname natexlab\endcsname\relax\def\natexlab#1{#1}\fi

\bibitem[{{Abdo} {et~al}\mbox{.}(2010){Abdo}, {Ackermann}, {Agudo}, {Ajello},
  {Aller}, {Aller}, {Angelakis}, {Arkharov}, {Axelsson}, {Bach}, \&
  et~al.}]{Abdo2010}
{Abdo} A.~A. {et~al.}, 2010, \apj, 716, 30

\bibitem[{{Abramowski} \& et~al. (HESS~Collaboration)(2013)}]{Aharonian13_0301}
{Abramowski} A., et~al. (HESS~Collaboration), 2013, \aap, 559, A136

\bibitem[{{Abramowski} \& et~al. (HESS~Collaboration)(2014)}]{Abramowski2014}
{Abramowski} A., et~al. (HESS~Collaboration), 2014, \aap, 571, A39

\bibitem[{{Aharonian} {et~al}\mbox{.}(2007){Aharonian}, {Akhperjanian},
  {Bazer-Bachi}, {Behera}, {Beilicke}, {Benbow}, {Berge}, {Bernl{\"o}hr},
  {Boisson}, {Bolz}, {Borrel}, {Boutelier}, {Braun}, {Brion}, {Brown},
  {B{\"u}hler}, {B{\"u}sching}, {Bulik}, {Carrigan}, {Chadwick}, {Clapson},
  {Chounet}, {Coignet}, {Cornils}, {Costamante}, {Degrange}, {Dickinson},
  {Djannati-Ata{\"i}}, {Domainko}, {Drury}, {Dubus}, {Dyks}, {Egberts},
  {Emmanoulopoulos}, {Espigat}, {Farnier}, {Feinstein}, {Fiasson},
  {F{\"o}rster}, {Fontaine}, {Funk}, {Funk}, {F{\"u}{\ss}ling}, {Gallant},
  {Giebels}, {Glicenstein}, {Gl{\"u}ck}, {Goret}, {Hadjichristidis}, {Hauser},
  {Hauser}, {Heinzelmann}, {Henri}, {Hermann}, {Hinton}, {Hoffmann}, {Hofmann},
  {Holleran}, {Hoppe}, {Horns}, {Jacholkowska}, {de Jager}, {Kendziorra},
  {Kerschhaggl}, {Kh{\'e}lifi}, {Komin}, {Kosack}, {Lamanna}, {Latham}, {Le
  Gallou}, {Lemi{\`e}re}, {Lemoine-Goumard}, {Lenain}, {Lohse}, {Martin},
  {Martineau-Huynh}, {Marcowith}, {Masterson}, {Maurin}, {McComb}, {Moderski},
  {Moulin}, {de Naurois}, {Nedbal}, {Nolan}, {Olive}, {Orford}, {Osborne},
  {Ostrowski}, {Panter}, {Pedaletti}, {Pelletier}, {Petrucci}, {Pita},
  {P{\"u}hlhofer}, {Punch}, {Ranchon}, {Raubenheimer}, {Raue}, {Rayner},
  {Renaud}, {Ripken}, {Rob}, {Rolland}, {Rosier-Lees}, {Rowell}, {Rudak},
  {Ruppel}, {Sahakian}, {Santangelo}, {Saug{\'e}}, {Schlenker}, {Schlickeiser},
  {Schr{\"o}der}, {Schwanke}, {Schwarzburg}, {Schwemmer}, {Shalchi}, {Sol},
  {Spangler}, {Stawarz}, {Steenkamp}, {Stegmann}, {Superina}, {Tam},
  {Tavernet}, {Terrier}, {van Eldik}, {Vasileiadis}, {Venter}, {Vialle},
  {Vincent}, {Vivier}, {V{\"o}lk}, {Volpe}, {Wagner}, {Ward}, \&
  {Zdziarski}}]{2155flare}
{Aharonian} F. {et~al.}, 2007, \apjl, 664, L71

\bibitem[{{Aliu} {et~al}\mbox{.}(2012){Aliu}, {Archambault}, {Arlen}, {Aune},
  {Beilicke}, {Benbow}, {B{\"o}ttcher}, {Bouvier}, {Bradbury}, {Buckley},
  {Bugaev}, {Byrum}, {Cannon}, {Cesarini}, {Ciupik}, {Collins-Hughes},
  {Connolly}, {Coppi}, {Cui}, {Decerprit}, {Dickherber}, {Dumm}, {Errando},
  {Falcone}, {Feng}, {Finley}, {Finnegan}, {Fortson}, {Furniss}, {Galante},
  {Gall}, {Godambe}, {Griffin}, {Grube}, {Gyuk}, {Hanna}, {Hawkins}, {Holder},
  {Huan}, {Hughes}, {Humensky}, {Kaaret}, {Karlsson}, {Kertzman}, {Khassen},
  {Kieda}, {Krawczynski}, {Krennrich}, {Lang}, {Lee}, {Madhavan}, {Maier},
  {Majumdar}, {McArthur}, {McCann}, {Moriarty}, {Mukherjee}, {Ong}, {Orr},
  {Otte}, {Palma}, {Park}, {Perkins}, {Pichel}, {Pohl}, {Prokoph}, {Quinn},
  {Ragan}, {Reyes}, {Reynolds}, {Roache}, {Rose}, {Ruppel}, {Saxon},
  {Schroedter}, {Sembroski}, {{\c S}ent{\"u}rk}, {Smith}, {Staszak},
  {Telezhinsky}, {Te{\v s}i{\'c}}, {Theiling}, {Thibadeau}, {Tsurusaki},
  {Varlotta}, {Vivier}, {Wakely}, {Ward}, {Weekes}, {Weinstein}, {Weisgarber},
  {Williams}, {Zitzer}, {Fortin}, \& {Horan}}]{Aliu12_0413}
{Aliu} E. {et~al.}, 2012, \apj, 750, 94

\bibitem[{{Bachev} {et~al}\mbox{.}(2015){Bachev}, {Spassov}, \&
  {Boeva}}]{ATel6944}
{Bachev} R., {Spassov} B., {Boeva} S., 2015, The Astronomer's Telegram, 6944, 1

\bibitem[{{Bachev} \& {Strigachev}(2015)}]{ATel6957}
{Bachev} R., {Strigachev} A., 2015, The Astronomer's Telegram, 6957, 1

\bibitem[{{Barthelmy} {et~al}\mbox{.}(2005){Barthelmy}, {Barbier}, {Cummings},
  {Fenimore}, {Gehrels}, {Hullinger}, {Krimm}, {Markwardt}, {Palmer},
  {Parsons}, {Sato}, {Suzuki}, {Takahashi}, {Tashiro}, \&
  {Tueller}}]{Barthelmy05}
{Barthelmy} S.~D. {et~al.}, 2005, \ssr, 120, 143

\bibitem[{{Begelman} {et~al}\mbox{.}(1984){Begelman}, {Blandford}, \&
  {Rees}}]{begelman84}
{Begelman} M.~C., {Blandford} R.~D., {Rees} M.~J., 1984, Reviews of Modern
  Physics, 56, 255

\bibitem[{Bevington \& Robinson(2003)}]{bevington2003data}
Bevington P., Robinson D., 2003, Data reduction and error analysis for the
  physical sciences, McGraw-Hill Higher Education. McGraw-Hill, Boston, MA

\bibitem[{{Bogdan} {et~al}\mbox{.}(2015){Bogdan}, {Alexandru}, \&
  {Raluca}}]{Bogdan15}
{Bogdan} D., {Alexandru} M., {Raluca} M.~G., 2015, Research in Astronomy and
  Astrophysics, 15, 327

\bibitem[{{B{\"o}ttcher} {et~al}\mbox{.}(2013){B{\"o}ttcher}, {Reimer},
  {Sweeney}, \& {Prakash}}]{Bottcher13}
{B{\"o}ttcher} M., {Reimer} A., {Sweeney} K., {Prakash} A., 2013, \apj, 768, 54

\bibitem[{{Burrows} {et~al}\mbox{.}(2005){Burrows}, {Hill}, {Nousek}, {Kennea},
  {Wells}, {Osborne}, {Abbey}, {Beardmore}, {Mukerjee}, {Short}, {Chincarini},
  {Campana}, {Citterio}, {Moretti}, {Pagani}, {Tagliaferri}, {Giommi},
  {Capalbi}, {Tamburelli}, {Angelini}, {Cusumano}, {Br{\"a}uninger}, {Burkert},
  \& {Hartner}}]{Burrows05}
{Burrows} D.~N. {et~al.}, 2005, \ssr, 120, 165

\bibitem[{{Cappi} {et~al}\mbox{.}(1994){Cappi}, {Comastri}, {Molendi},
  {Palumbo}, {Della Ceca}, \& {Maccacaro}}]{Cappi94}
{Cappi} M., {Comastri} A., {Molendi} S., {Palumbo} G.~G.~C., {Della Ceca} R.,
  {Maccacaro} T., 1994, \mnras, 271, 438

\bibitem[{{Carrasco} {et~al}\mbox{.}(2015){Carrasco}, {Porras}, {Recillas},
  {Leon-Tavares}, {Chavushyan}, \& {Carraminana}}]{ATel6902}
{Carrasco} L., {Porras} A., {Recillas} E., {Leon-Tavares} J., {Chavushyan} V.,
  {Carraminana} A., 2015, The Astronomer's Telegram, 6902, 1

\bibitem[{{Dai} {et~al}\mbox{.}(2013){Dai}, {Wu}, {Zhu}, {Zhou}, {Ma}, {Yuan},
  \& {Wang}}]{Dai13}
{Dai} Y., {Wu} J., {Zhu} Z.-H., {Zhou} X., {Ma} J., {Yuan} Q., {Wang} L., 2013,
  \apjs, 204, 22

\bibitem[{{Donato} {et~al}\mbox{.}(2001){Donato}, {Ghisellini}, {Tagliaferri},
  \& {Fossati}}]{Donato01}
{Donato} D., {Ghisellini} G., {Tagliaferri} G., {Fossati} G., 2001, \aap, 375,
  739

\bibitem[{{Donato} {et~al}\mbox{.}(2005){Donato}, {Sambruna}, \&
  {Gliozzi}}]{Donato05}
{Donato} D., {Sambruna} R.~M., {Gliozzi} M., 2005, \aap, 433, 1163

\bibitem[{{Ferrero} {et~al}\mbox{.}(2006){Ferrero}, {Wagner},
  {Emmanoulopoulos}, \& {Ostorero}}]{Ferrero06}
{Ferrero} E., {Wagner} S.~J., {Emmanoulopoulos} D., {Ostorero} L., 2006, \aap,
  457, 133

\bibitem[{{Foschini} {et~al}\mbox{.}(2006){Foschini}, {Tagliaferri}, {Pian},
  {Ghisellini}, {Treves}, {Maraschi}, {Tavecchio}, {Di Cocco}, \&
  {Rosen}}]{Foschini06}
{Foschini} L. {et~al.}, 2006, \aap, 455, 871

\bibitem[{{Gehrels} {et~al}\mbox{.}(2004){Gehrels}, {Chincarini}, {Giommi},
  {Mason}, {Nousek}, {Wells}, {White}, {Barthelmy}, {Burrows}, {Cominsky},
  {Hurley}, {Marshall}, {M{\'e}sz{\'a}ros}, {Roming}, {Angelini}, {Barbier},
  {Belloni}, {Campana}, {Caraveo}, {Chester}, {Citterio}, {Cline}, {Cropper},
  {Cummings}, {Dean}, {Feigelson}, {Fenimore}, {Frail}, {Fruchter}, {Garmire},
  {Gendreau}, {Ghisellini}, {Greiner}, {Hill}, {Hunsberger}, {Krimm},
  {Kulkarni}, {Kumar}, {Lebrun}, {Lloyd-Ronning}, {Markwardt}, {Mattson},
  {Mushotzky}, {Norris}, {Osborne}, {Paczynski}, {Palmer}, {Park}, {Parsons},
  {Paul}, {Rees}, {Reynolds}, {Rhoads}, {Sasseen}, {Schaefer}, {Short},
  {Smale}, {Smith}, {Stella}, {Tagliaferri}, {Takahashi}, {Tashiro},
  {Townsley}, {Tueller}, {Turner}, {Vietri}, {Voges}, {Ward}, {Willingale},
  {Zerbi}, \& {Zhang}}]{Gehrels04}
{Gehrels} N. {et~al.}, 2004, \apj, 611, 1005

\bibitem[{{Giommi} {et~al}\mbox{.}(1999){Giommi}, {Massaro}, {Chiappetti},
  {Ferrara}, {Ghisellini}, {Jang}, {Maesano}, {Miller}, {Montagni}, {Nesci},
  {Padovani}, {Perlman}, {Raiteri}, {Sclavi}, {Tagliaferri}, {Tosti}, \&
  {Villata}}]{Giommi99}
{Giommi} P. {et~al.}, 1999, \aap, 351, 59

\bibitem[{{Gopal-Krishna} {et~al}\mbox{.}(2011){Gopal-Krishna}, {Goyal},
  {Joshi}, {Karthick}, {Sagar}, {Wiita}, {Anupama}, \&
  {Sahu}}]{Gopal-Krishna11}
{Gopal-Krishna}, {Goyal} A., {Joshi} S., {Karthick} C., {Sagar} R., {Wiita}
  P.~J., {Anupama} G.~C., {Sahu} D.~K., 2011, \mnras, 416, 101

\bibitem[{{Kalberla} {et~al}\mbox{.}(2005){Kalberla}, {Burton}, {Hartmann},
  {Arnal}, {Bajaja}, {Morras}, \& {P{\"o}ppel}}]{Kalberla05}
{Kalberla} P.~M.~W., {Burton} W.~B., {Hartmann} D., {Arnal} E.~M., {Bajaja} E.,
  {Morras} R., {P{\"o}ppel} W.~G.~L., 2005, \aap, 440, 775

\bibitem[{{Liao} \& {Bai}(2015)}]{Liao15}
{Liao} N.~H., {Bai} J.~M., 2015, \na, 34, 134

\bibitem[{{Liao} {et~al}\mbox{.}(2014){Liao}, {Bai}, {Liu}, {Weng}, {Chen}, \&
  {Li}}]{Liao14}
{Liao} N.~H., {Bai} J.~M., {Liu} H.~T., {Weng} S.~S., {Chen} L., {Li} F., 2014,
  \apj, 783, 83

\bibitem[{{Mirzoyan}(2015)}]{ATel6999}
{Mirzoyan} R., 2015, The Astronomer's Telegram, 6999, 1

\bibitem[{{Montagni} {et~al}\mbox{.}(2006){Montagni}, {Maselli}, {Massaro},
  {Nesci}, {Sclavi}, \& {Maesano}}]{Montagni06}
{Montagni} F., {Maselli} A., {Massaro} E., {Nesci} R., {Sclavi} S., {Maesano}
  M., 2006, \aap, 451, 435

\bibitem[{{Padovani} \& {Giommi}(1996)}]{Padovani96}
{Padovani} P., {Giommi} P., 1996, \mnras, 279, 526

\bibitem[{{Roming} {et~al}\mbox{.}(2005){Roming}, {Kennedy}, {Mason}, {Nousek},
  {Ahr}, {Bingham}, {Broos}, {Carter}, {Hancock}, {Huckle}, {Hunsberger},
  {Kawakami}, {Killough}, {Koch}, {McLelland}, {Smith}, {Smith}, {Soto},
  {Boyd}, {Breeveld}, {Holland}, {Ivanushkina}, {Pryzby}, {Still}, \&
  {Stock}}]{Roming05}
{Roming} P.~W.~A. {et~al.}, 2005, \ssr, 120, 95

\bibitem[{{Saito} {et~al}\mbox{.}(2013){Saito}, {Stawarz}, {Tanaka},
  {Takahashi}, {Madejski}, \& {D'Ammando}}]{saito}
{Saito} S., {Stawarz} {\L}., {Tanaka} Y.~T., {Takahashi} T., {Madejski} G.,
  {D'Ammando} F., 2013, \apjl, 766, L11

\bibitem[{{Spiridonova} {et~al}\mbox{.}(2015){Spiridonova}, {Vlasyuk},
  {Moskvitin}, \& {Bychkova}}]{ATel7004}
{Spiridonova} O.~I., {Vlasyuk} V.~V., {Moskvitin} A.~S., {Bychkova} V.~S.,
  2015, The Astronomer's Telegram, 7004, 1

\bibitem[{{Tagliaferri} {et~al}\mbox{.}(2003){Tagliaferri}, {Ravasio},
  {Ghisellini}, {Giommi}, {Massaro}, {Nesci}, {Tosti}, {Aller}, {Aller},
  {Celotti}, {Maraschi}, {Tavecchio}, \& {Wolter}}]{Tagliaferri03}
{Tagliaferri} G. {et~al.}, 2003, \aap, 400, 477

\bibitem[{{Takahashi} {et~al}\mbox{.}(1999){Takahashi}, {Madejski}, \&
  {Kubo}}]{Takahashi99}
{Takahashi} T., {Madejski} G., {Kubo} H., 1999, Astroparticle Physics, 11, 177

\bibitem[{{Takahashi} {et~al}\mbox{.}(1996){Takahashi}, {Tashiro}, {Madejski},
  {Kubo}, {Kamae}, {Kataoka}, {Kii}, {Makino}, {Makishima}, \&
  {Yamasaki}}]{Takahashi96}
{Takahashi} T. {et~al.}, 1996, \apjl, 470, L89

\bibitem[{{Vercellone} {et~al}\mbox{.}(2011){Vercellone}, {Striani},
  {Vittorini}, {Donnarumma}, {Pacciani}, {Pucella}, {Tavani}, {Raiteri},
  {Villata}, {Romano}, {Fiocchi}, {Bazzano}, {Bianchin}, {Ferrigno},
  {Maraschi}, {Pian}, {T{\"u}rler}, {Ubertini}, {Bulgarelli}, {Chen},
  {Giuliani}, {Longo}, {Barbiellini}, {Cardillo}, {Cattaneo}, {Del Monte},
  {Evangelista}, {Feroci}, {Ferrari}, {Fuschino}, {Gianotti}, {Giusti},
  {Lazzarotto}, {Pellizzoni}, {Piano}, {Pilia}, {Rapisarda}, {Rappoldi},
  {Sabatini}, {Soffitta}, {Trifoglio}, {Trois}, {Giommi}, {Lucarelli},
  {Pittori}, {Santolamazza}, {Verrecchia}, {Agudo}, {Aller}, {Aller},
  {Arkharov}, {Bach}, {Berdyugin}, {Borman}, {Chigladze}, {Efimov}, {Efimova},
  {G{\'o}mez}, {Gurwell}, {McHardy}, {Joshi}, {Kimeridze}, {Krajci},
  {Kurtanidze}, {Kurtanidze}, {Larionov}, {Lindfors}, {Molina}, {Morozova},
  {Nazarov}, {Nikolashvili}, {Nilsson}, {Pasanen}, {Reinthal}, {Ros}, {Sadun},
  {Sakamoto}, {Sallum}, {Sergeev}, {Schwartz}, {Sigua}, {Sillanp{\"a}{\"a}},
  {Sokolovsky}, {Strelnitski}, {Takalo}, {Taylor}, \& {Walker}}]{Vercellone}
{Vercellone} S. {et~al.}, 2011, \apjl, 736, L38

\bibitem[{{Wagner}(2009)}]{Wagner2009}
{Wagner} S., 2009, in Astrophysics with All-Sky X-Ray Observations. RIKEN, and
  JAXA Suzuki Umetaro Hall, RIKEN Wako, Saitama, Japan, {Kawai} N., {Mihara}
  T., {Kohama} M., {Suzuki} M., eds., p. 186

\bibitem[{{Wagner}(1992)}]{Wagner92}
{Wagner} S.~J., 1992, in X-ray Emission from Active Galactic Nuclei and the
  Cosmic X-ray Background, Max-Planck-Institut fur Extraterrestrische Physik,
  Garching bei Munchen, Germany, {Brinkmann} W., {Truemper} J., eds., pp.
  97--102

\bibitem[{{Wagner} \& {Witzel}(1995)}]{wagner}
{Wagner} S.~J., {Witzel} A., 1995, \araa, 33, 163

\bibitem[{{Wu} {et~al}\mbox{.}(2014){Wu}, {Dai}, {Zhou}, \& {Ma}}]{Wu14}
{Wu} J., {Dai} Y., {Zhou} X., {Ma} J., 2014, JA\&A

\end{thebibliography}

\label{lastpage}

\end{document}